\documentclass{basi}
\usepackage[T1]{fontenc}
\usepackage[british]{babel}
\usepackage[varg]{txfonts}
\usepackage{rotating}
\usepackage{dcolumn}
\usepackage{graphicx}

\begin{document}

\title[X-ray/optical view of Type-I bursts in EXO 0748-676]{Simultaneous X-ray and optical observations of thermonuclear bursts in the LMXB EXO 0748-676}

\author[B. Paul et al.]{Biswajit Paul\thanks{E-mail:
bpaul@rri.res.in (BP); archanam79@gmail.com (AM), lsaripal@rri.res.in (LS)}, Archana M. \& Lakshmi Saripalli\\
Raman Research Institute, Sadashivnagar, C. V. Raman Avenue, Bangalore 560080, India}

\pubyear{2012}
\volume{00}
\pagerange{\pageref{firstpage}--\pageref{lastpage}}

\date{Received --- ; accepted ---}

\maketitle

\label{firstpage}

\begin{abstract}

We report the detection of a large number of optical bursts in the Low Mass X-ray Binary (LMXB) EXO 0748-676 simultaneous with the thermonuclear X-ray bursts. The X-ray and the optical bursts are detected in a long observation of this source with the XMM-Newton observatory. This has increased the number of thermonuclear X-ray bursts in the LMXBs with simultaneous optical detection by several factors. The optical bursts are found to have a linear rise followed by a slow, somewhat exponential decay. Most of the optical bursts have longer rise and decay timescale compared to the corresponding X-ray bursts. We have determined the X-ray and optical excess photon counts in the bursts that allow us to look at the optical to X-ray burst fluence ratio for each burst and the ratio as a function of the X-ray burst intensity and as a function of the orbital phase. The delay between the onset of the X-ray bursts and the onset of the optical bursts have also been measured and is found to have an average value of 3.25 seconds. We do not find any convincing evidence of orbital phase dependence of the following parameters: X-ray to optical delay, rise time of the optical bursts, and optical to X-ray burst intensity ratio as would be expected if the optical bursts were produced by reprocessing from the surface of the secondary star that is facing the compact star. On the other hand, if the optical bursts are produced by reprocessing of the X-rays in the accretion disk, the onset of the bursts is not expected to have a sharp, linear shape as is observed in a few of the bursts in EXO 0748-676. We emphasise the fact that simultaneous optical observations of the X-ray bursts in multiple wavelength bands will enable further detailed investigations of the reprocessing phenomena, including any non-linear effect of the X-ray irradiation.

\end{abstract}


\section{Introduction}

The Low Mass X-ray Binary (LMXB) EXO 0748-67 was discovered in 1985 (Parmar et al. 1986) as a transient X-ray 
source and its optical counterpart, UY Vol was discovered in the same year (Wade et al. 1985). EXO 0748-676 is a 
well known LMXB that has been the target of detailed timing and spectroscopic observations ever since its discovery. 
Unlike most X-ray transients, it did not decay back to quiescent state, but remained active and was  considered part 
of the persistent LMXB population. In late 2008 it started a transition to quiescence after it had been actively 
accreting for more than 24 years (Hynes \& Jones 2009). During 1985 to 2008 it was a moderately bright LMXB with strong dipping 
activities. It is one of the rare LMXBs  that show a complete X-ray eclipse, every 3.82 h. Dips and eclipses are 
due to the central X-ray source being obscured by some structure above the disk and occulted by the companion star 
respectively at every orbital period. The presence of eclipses and complex dipping activity indicates that the 
object is viewed at an angle of inclination in the range of 75$^\circ$-82$^\circ$ (very close to the accretion disc 
plane) and the mass of the companion is constrained to be about 0.5 solar mass (Parmar et al. 1986). The compact source 
is known to be a weakly magnetized neutron star because of the thermonuclear burst activities. Two burst oscillation 
features at 45 Hz (Villarreal \& Strohmayer 2004) and 550 Hz (Galloway et al. 2010) were detected in this LMXB 
but Jain \& Paul (2011) reported non detection of any oscillations around 45 Hz down to a very small pulse fraction. 
The distance to EXO 0748-676 was derived as 7.7 kpc for a helium dominated burst photosphere and 5.9 kpc for a 
hydrogen dominated burst photosphere and a strong X-ray burst from the same source is an evidence for photospheric 
expansion (Wolff et al. 2005). Another unusual feature of EXO 0748-676 is presence of orbital period glitches (Wolff et al. 
2009); only one other such system is known, XTE J1710-281 (Jain et al. 2011).
It is also found to show a rich variety of burst profiles and also some double and triple X-ray bursts (Boirin et al. 
2007).

A consequence of the thermonuclear X-ray bursts observed in the low magnetic field accreting neutron stars
(apart from being directly received by the observer) is 
reprocessing of the X-ray photons as they interact with different components of the binary, mainly the companion 
star and the accretion disk. The reprocessed X-rays, often seen at optical wavelengths are delayed in time with 
respect to the X-ray bursts due to light travel time and reprocessing effects. Simultaneous X-ray and optical 
bursts have been detected in several LMXBs:
Ser X-1 (Hackwell et al. 1979),
4U 1536-53 (Pederson et al. 1982; Lawrence et al. 1983; Matsuoka et al. 1984 ),
GS 1826-24 (Kong et al. 2000),
MS 1603+2600 (Hakala et al. 2005) and
EXO 0748-676 (Hynes et al. 2006).
Simultaneous X-ray and UV/optical
observations offer a tremendous opportunity to probe the geometry and physics of the 
X-ray binaries.
EXO 0748-676 is one of the very few eclipsing LMXBs that allows us to investigate 
the reprocessing phenomena as a function of orbital phase. If the X-rays are reprocessed from the surface of the 
companion star that faces the X-ray star, one expects very strong orbital phase dependence of the reprocessing 
parameters that can be tested well in the case of EXO 0748-676.

The Optical Monitor on the XMM-Newton observatory allows examination of the optical and ultra-violet lightcurves
simultaneously alongside an X-ray burst. This adds an important dimension to the study of LMXBs where 
one can relate the optical emission to the X-ray bursts and hence study the reprocessing mechanisms and examine 
the orbital characteristics, structure of the binary and the accretion disk. Several LMXBs have been observed with 
XMM with very long exposures and data from the optical monitor of XMM should be useful in investigating the optical 
reprocessing of the X-ray bursts.
Compared to the literature on the thermonuclear X-ray bursts, work based on simultaneous observation of X-ray and optical bursts is quite limited owing to the difficulties in carrying out optical observations (which are mostly done from ground based observatories) simultaneously with X-ray observations (done from space).
Here we report on our examination of the archival data from simultaneous X-ray and optical monitoring of 
EXO 0748-676 with the XMM-Newton Observatory.

\section{Observations, analysis and results}

The source was observed with the XMM-Newton observatory (Jansen et al. 2001) on several occasions with different instrument 
configurations. XMM has 3 X-ray mirrors with differemt focal plane CCD (2 MOS CCDS and 1 PN CCD) imaging 
spectrometers that cover the energy band of 0.1 - 12 keV with a collecting area of 1900 cm$^2$ at 150 eV that 
decreases to 350 cm$^2$ at 10 keV for each of the the telescopes (Struder et al. 2001, Turner et al. 2001).
76 X-ray bursts were recorded during the 7 XMM-newton observations of EXO 0748-676 during September-November 2003 
(Boirin et al. 2007). The 
XMM Optical Monitor (XMM-OM) instrument (Mason et al. 2001) provides coverage between 180 nm and 600 nm of the 
central 17 arc minute square region of the X-ray field of view, permitting multi wavelength observations of XMM 
targets simultaneously in the X-ray and ultraviolet/optical bands.
The simultaneous optical observations were performed with the XMM-OM in white light. 

Though 
until 2008, EXO 0748-676 was considered to be a persistent LMXB, it had strong long term intenisty variations and 
the source was bright during these XMM observations.
Data reduction and analysis were carried out with XMM Science Analysis Software (SAS). We processed the data using
the tasks omfchain and epchain.  We extracted optical light curves 
from a circular region of radius 6 arcsec (FWHM) around the optical counterpart while the X-ray light curves 
were extracted from the EPIC-PN CCD from a region of radius 11 arcsec around the source position. Both the 
light curves had a bin size of 1 s. All the X-ray bursts reported earlier (Boirin et al. 2007) were detected in the 
energy band of 0.2-12 keV and also in two subbands of 0.2-2.5 keV and 5-12 keV. Apart from the bursts, the low 
energy light curve also shows significant orbital modulation with significant variation from orbit to orbit. 
The 5-12 keV light curve did not show any significant variation apart form the bursts. 63 of the X-ray bursts had 
simultaneous optical data and all but one (burst no 52) of these bursts were also detected in the optical light curves. For 13 of the X-ray 
bursts, there was no simultaneous optical data. The profiles of the optical bursts are shown in Figure 1 along with 
the X-ray bursts. The bursts have been numbered in the same way as in Boirin et al. (2007).
A few selected bursts are shown separately in Figure 2, to clearly demonstrate some of the burst characteristics.

Some important features noticed in the light curves are as follows.
\begin{itemize}
\item Both the X-ray and the optical bursts have linear, fast rise and slow, somewhat exponential decay.
\item The start of the optical bursts have a delay compared to the X-ray bursts.
\item The optical bursts are generally of longer duration.
\item Weak correlation between the peak amplitude of the X-ray and optical bursts.
\item Some of the burst profiles show clear deviations from an exponential shape. 
\end{itemize}

Echo, or reverberation mapping technique has been applied extensively on the multi-wavelength profiles of thermonuclear bursts in EXO 0748-676 (Hynes et al. 2006).
This assumes an instantaneous reprocessing of the incident X-ray flux into UV and optical light, the transfer function calculated for each brust representing the weighted geometric delays from different parts of the reprocessing region. However, for the thermonuclear bursts in EXO 0748-676, and probably for other sources as well, there is strong energy dependence of the burst profiles. The decay time scales of bursts increase from hard X-rays to soft X-rays, UV, and optical. The transfer function often extends upto 10 seconds or more, considerably larger than the light travel time accross the X-ray binary. These are indications of a complex reprocessing and temperature evolution of the reprocessed component during the bursts. Considering the temperature evolution that is seen during the thermonuclear X-ray bursts, a temperature evolution of the reprocessed emission during the bursts is certainly possible. Moreover, rather than the X-ray count rate, the reprocessed emission should be related to the X-ray flux rate.
In view of these complexities and the fact that the optical light curve in our dataset corresponds only to a single broad band, and also that the  X-ray count rate is not large enough to determine the flux/temperature evolution accurately, we chose to limit the scope of this work.
In the present work therefore, instead of calculating the transfer functions, we determine the burst start times, rise times, and total burst counts in the X-ray and optical bands.

We fitted all the burst profiles with a model consisting of a constant component, a burst with a linear rise and an exponential decay. As mentioned earlier, some of the burst profiles are not fitted well with a single exponential decay. However, the model used is sufficient to investigate some important features of the X-ray bursts.
From this model fitting we were able to derive the constant emission component before the bursts, start time, peak time and the exponential decay time scale. The X-ray and optical counts of the bursts were obtained by subtracting the pre-burst component from the total photon counts during the bursts.
The orbital phase of the start time of each burst was calculated using an orbital period of 0.15933775443 day and reference mid-eclipse time of 52903.963579 MJD that is appropriate for this epoch (Wolff et al. 2009).

In the left panel of Figure 3 we have shown a plot of the optical counts of the 62 bursts detected in the XMM-OM light curves against the X-ray counts (0.2-12 keV) while the ratio of the optical counts to  the X-ray counts for each burst is shown against the X-ray counts in the right panel of the same figure. Henceforth, we call the ratio of the optical counts to the X-ray counts for each burst as optical conversion factor.
A correlation is seen between the X-ray and the optical counts of the bursts (shown with a straight line) although this is accompanied by a large scatter.
In the three panels of Figure 4, we have shown the delay between the start time of the X-ray and the optical bursts, the risetime of the optical bursts and the optical conversion factor as function of the orbital phases of the bursts, with the mid-eclipse time represented as phase zero. The lines plotted through the filled circles in each panel represent the average value of the parameters in wider orbital phase intervals of 0.1. The delay between the X-ray and the optical bursts has a maximum of about 8 seconds with an average value of 3.25 seconds.
However for none of the three parameters, X-ray to optical delay, optical risetime, or optical conversion factor, is there a clear orbital phase dependence seen that can be expected for reprocessing from the surface of the companion star .

\section{Discussion}

In the LMXBs, a large fraction of the optical emission can arise from X-rays reprocessed by the material in regions 
around the central compact object. Search for this reprocessed variability requires very long simultaneous X-ray and 
optical observations, which is usually difficult to carry out. The thermonuclear X-ray bursts (Type-1 X-ray bursts) 
in LMXBs involve a large increase in the X-ray flux, by a factor of 10 or more on timescales of a few seconds. As 
well as providing insights into the conditions on the surface of the neutron star, the sudden flash lights up the 
whole binary system, and they are expected to be manifested in the UV and optical via reprocessed X-ray emission. The 
optical bursts almost certainly arise as a consequence of the reprocessing of a fraction of the X-ray burst energy 
by matter located within a few light seconds of the neutron star. The accretion disk around the neutron star and the 
surface of the companion star are plausible sites for the reprocessing. The Type-1 X-ray bursts serve as a probe to 
obtain information about the geometry of the X-ray binary by illuminating the surroundings of the compact object. 
The reprocessing matter absorbs part of the X-rays, get heated and re-radiates the absorbed energy with an energy 
distribution related to its temperature. By comparing the features of the X-ray bursts that originate on the compact 
star, with the corresponding features of optical bursts that originate from the surroundings, one can obtain 
information about the surroundings (Pedersen et al. 1982). The optical burst may contain only a fraction of the flux 
of the X-ray bursts depending on the solid angle that the reprocessor subtends on the compact start and the 
reprocessing efficiency. From analysis of the simultaneous X-ray and optical light curves presented here it is clear 
that all the X-ray bursts are accompanied by optical bursts. The resulting reprocessed variability will be delayed 
in time with respect to the X-ray variability by an amount depending on the position of the reprocessing regions 
which in turn depends on the geometry of the binary and its orientation with respect to the observer's line of sight. The reprocessed optical emission seen by a distant observer will be delayed in time due to a combination of light travel time differences and radiation reprocessing time. One may expect the optical radiative delays to be much smaller than the light travel time differences (Pedersen et al. 1982). But most, if not all of the optical bursts are seen to be continued several seconds after the X-ray burst ended (Figure 1 \& Figure 2). These observations indicate that there is a radiative delay and radiative smearing of the optical flux relative to the X-ray flux.

In the present work, we have used the rising parts of the X-ray and optical bursts which contain clear information 
related to the light travel time differences and thus about the  geometry of the reprocessing region. As different 
parts of the reprocessing body give rise to different values of this geometric delay there will be a geometric 
smearing also. The reprocessed radiation can be considered as a superposition of contributions from different parts 
of the reprocessing body.
If the X-ray absorbing material in the disc  has a significant scale height above the mid-plane so that the 
companion would be effectively shielded from direct X-ray illumination, it will reduce the strength of reprocessing 
from the surface of the companion star.

In agreement with the reprocessing scenario, the optical bursts were observed a few seconds after the X-ray bursts. 
Average value of the delay is 3.25 seconds. The light travel time accross this compact X-ray binary system is 
about 3 seconds (Hynes et al. 2006) and therefore, the average delay of the optical bursts with respect to the X-ray 
bursts is compatible with the reprocessing region being the surface of the compact start or the outer accretion disk.

It has been proposed that the optical reprocessing takes place from the surface of the companion star facing the 
compact star and the accretion disk. Investigations have been made for EXO 0748-676 using simultaneous X-ray and 
multi band optical/UV observations (Hynes et al. 2006). In the present work, apart from detection of a large number 
of simultaneous X-ray and optical bursts, an important outcome is that we did not find any convincing evidence of 
orbital phase dependence of the following parameters: X-ray to optical delay, rise time of the optical bursts, and 
optical conversion factor as would be expected if the optical bursts were produced by reprocessing from the surface 
of the secondary star that is facing the compact star. In particular, the optical conversion factor is expected to 
be very small if the orbital phase is near zero. On the other hand, if the optical bursts are produced by 
reprocessing of the X-rays in the accretion disk, the onset of the bursts is not expected to have a sharp, linear 
shape as is observed in a few of the bursts in EXO 0748-676.

The temperature and amplitude of the reprocessed emission is expected to evolve during the bursts and is also 
expected to be different in different bursts. Thus observations carried out in a single optical/UV band is expected 
to show a complex non-linear relation between the counts in the X-ray and the optical bursts. In absence of multi 
band optical/UV measurement, we refrain from any discussion about the correlation between the X-ray and the optical 
burst amplitudes except for examining the simple optical conversion factor as a funtion of the orbital phase of the 
binary.

The average X-ray to optical flux ratio of the bursts is calculated to be about 300 with some uncertainity due to our lack of knowledge about the optical spectral characetristics of the bursts. It is evident from Figures 3 \& 4 that there is a large difference from burst to burst by a factor of as much as five. Accurate determination of this ratio from multi band optical/UV observations could be useful for determining the fraction of X-rays which are intercepted by the disk which will in turn be useful for estimation of the thickness of the disk as seen from the neutron star. The upcoming multi-wavelength observatory ASTROSAT with large area X-ray detectors (Paul 2009) and Optical-UV telescopes (Srivastava et al. 2009) will be of particular interest for such studies.

\section{Conclusions}

We have detected a large number of simultaneous X-ray and optical Type-I bursts in the eclipsing LMXB EXO 0748-676
and have investigated the delay of the optical bursts, the optical conversion factor etc. No significant orbital 
phase dependence is seen for these parameters which argues against the surface of the companion star being the 
reprocessing region. We emphasise the fact that simultaneous optical observations of the X-ray bursts in multiple 
wavelength bands will enable further detailed investigations of the reprocessing phenomena, including any non-linear 
effect of the X-ray irradiation.

\section{Acknowledgement}
This research has made use of data obtained from the High Energy Astrophysics Science Archive Research Center 
(HEASARC), provided by the NASA Goddard Space Flight Center.

\clearpage

\begin{figure}
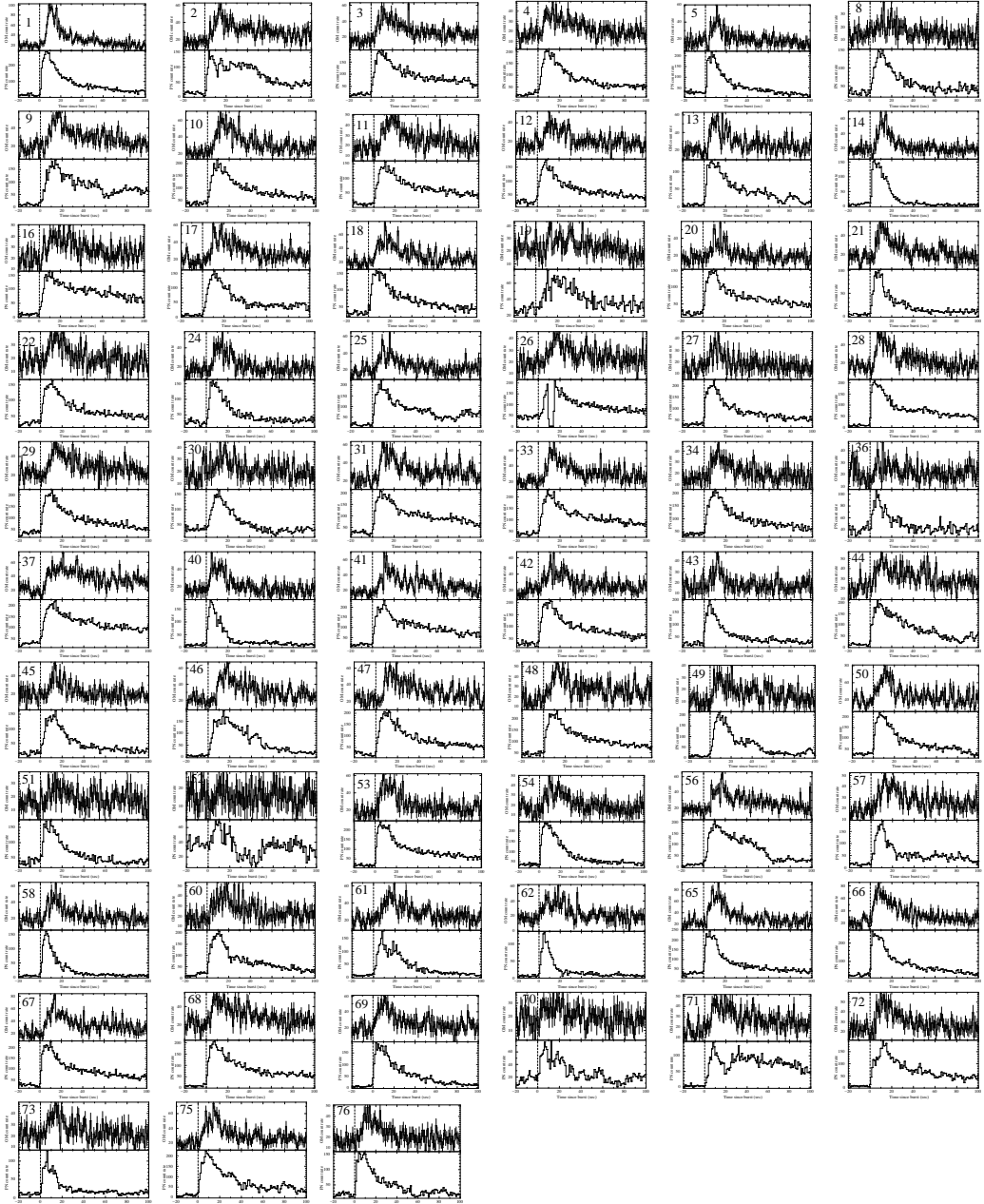

\includegraphics[width=15mm,angle=-90]{1.ps}
\includegraphics[width=15mm,angle=-90]{2.ps}
\includegraphics[width=15mm,angle=-90]{3.ps}
\includegraphics[width=15mm,angle=-90]{4.ps}
\includegraphics[width=15mm,angle=-90]{5.ps}
\includegraphics[width=15mm,angle=-90]{8.ps}
\includegraphics[width=15mm,angle=-90]{9.ps}
\includegraphics[width=15mm,angle=-90]{10.ps}
\includegraphics[width=15mm,angle=-90]{11.ps}
\includegraphics[width=15mm,angle=-90]{12.ps}
\includegraphics[width=15mm,angle=-90]{13.ps}
\includegraphics[width=15mm,angle=-90]{14.ps}
\includegraphics[width=15mm,angle=-90]{16.ps}
\includegraphics[width=15mm,angle=-90]{17.ps}
\includegraphics[width=15mm,angle=-90]{18.ps}
\includegraphics[width=15mm,angle=-90]{19.ps}
\includegraphics[width=15mm,angle=-90]{20.ps}
\includegraphics[width=15mm,angle=-90]{21.ps}
\includegraphics[width=15mm,angle=-90]{22.ps}
\includegraphics[width=15mm,angle=-90]{24.ps}
\includegraphics[width=15mm,angle=-90]{25.ps}
\includegraphics[width=15mm,angle=-90]{26.ps}
\includegraphics[width=15mm,angle=-90]{27.ps}
\includegraphics[width=15mm,angle=-90]{28.ps}
\includegraphics[width=15mm,angle=-90]{29.ps}
\includegraphics[width=15mm,angle=-90]{30.ps}
\includegraphics[width=15mm,angle=-90]{31.ps}
\includegraphics[width=15mm,angle=-90]{33.ps}
\includegraphics[width=15mm,angle=-90]{34.ps}
\includegraphics[width=15mm,angle=-90]{36.ps}
\includegraphics[width=15mm,angle=-90]{37.ps}
\includegraphics[width=15mm,angle=-90]{40.ps}
\includegraphics[width=15mm,angle=-90]{41.ps}
\includegraphics[width=15mm,angle=-90]{42.ps}
\includegraphics[width=15mm,angle=-90]{43.ps}
\includegraphics[width=15mm,angle=-90]{44.ps}
\includegraphics[width=15mm,angle=-90]{45.ps}
\includegraphics[width=15mm,angle=-90]{46.ps}
\includegraphics[width=15mm,angle=-90]{47.ps}
\includegraphics[width=15mm,angle=-90]{48.ps}
\includegraphics[width=15mm,angle=-90]{49.ps}
\includegraphics[width=15mm,angle=-90]{50.ps}
\includegraphics[width=15mm,angle=-90]{51.ps}
\includegraphics[width=15mm,angle=-90]{52.ps}
\includegraphics[width=15mm,angle=-90]{53.ps}
\includegraphics[width=15mm,angle=-90]{54.ps}
\includegraphics[width=15mm,angle=-90]{56.ps}
\includegraphics[width=15mm,angle=-90]{57.ps}
\includegraphics[width=15mm,angle=-90]{58.ps}
\includegraphics[width=15mm,angle=-90]{60.ps}
\includegraphics[width=15mm,angle=-90]{61.ps}
\includegraphics[width=15mm,angle=-90]{62.ps}
\includegraphics[width=15mm,angle=-90]{65.ps}
\includegraphics[width=15mm,angle=-90]{66.ps}
\includegraphics[width=15mm,angle=-90]{67.ps}
\includegraphics[width=15mm,angle=-90]{68.ps}
\includegraphics[width=15mm,angle=-90]{69.ps}
\includegraphics[width=15mm,angle=-90]{70.ps}
\includegraphics[width=15mm,angle=-90]{71.ps}
\includegraphics[width=15mm,angle=-90]{72.ps}
\includegraphics[width=15mm,angle=-90]{73.ps}
\includegraphics[width=15mm,angle=-90]{75.ps}
\includegraphics[width=15mm,angle=-90]{76.ps}
\caption{The V band optical light curve and the 0.2-12 keV X-ray lightcurve of each of the thermonuclear bursts are shown here at the top and bottom of each panel. The X-axis range is identical for all the bursts, and includes 20 seconds before and 100 seconds after the onset of the X-ray bursts.}
\end{figure}

\begin{figure}
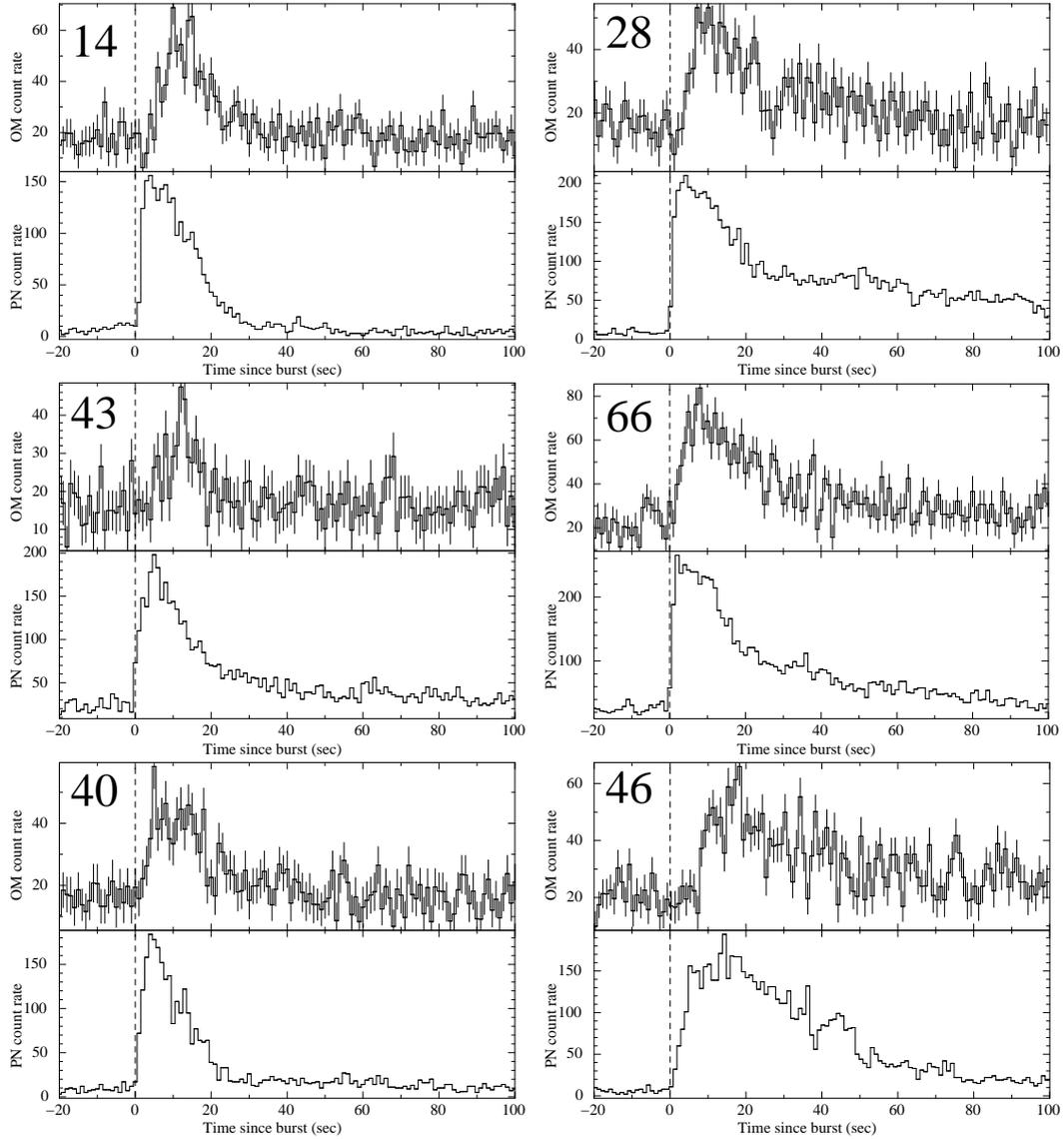

\includegraphics[width=50mm,angle=-90]{14.ps}
\includegraphics[width=50mm,angle=-90]{28.ps}
\includegraphics[width=50mm,angle=-90]{43.ps}
\includegraphics[width=50mm,angle=-90]{66.ps}
\includegraphics[width=50mm,angle=-90]{40.ps}
\includegraphics[width=50mm,angle=-90]{46.ps}
\caption{A few of the light curves shown in Figure 1 are shown here for more clarity with some characteristic features. No 14 is a burst that has an exponential decay; No 28 is a burst that significantly differes from an exponential decay; No 43 is a burst with a weak optical component. The three bursts 66, 40 and 46 are shown here have small, medium and large X-ray to optical delay respectively. The burst No 46 also has a very small rise time in the optical band.} 
\end{figure}

\clearpage

\begin{figure}
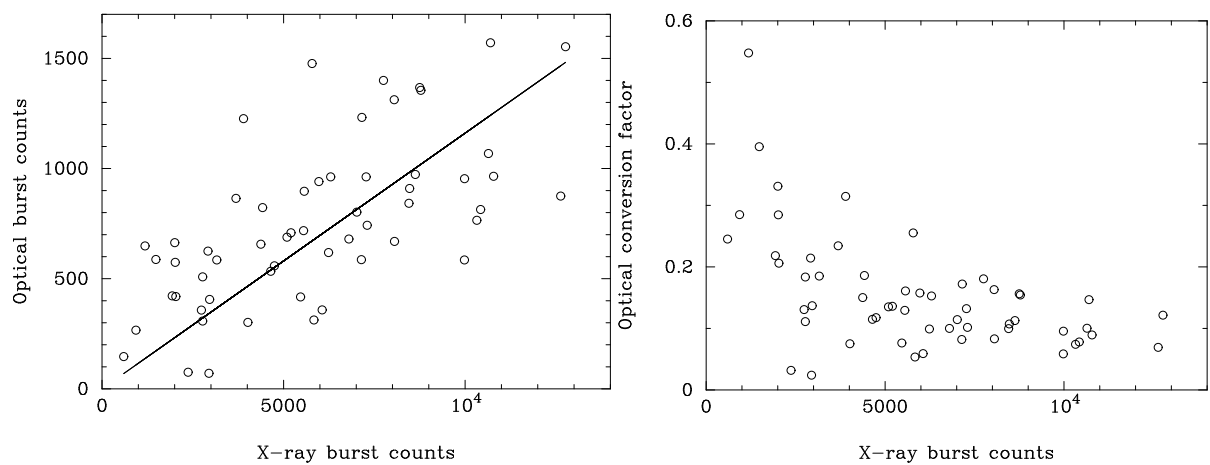

\includegraphics[width=60mm,angle=-90]{lopt_lx.ps}
\vspace{2cm}
\includegraphics[width=60mm,angle=-90]{ratio_lx.ps}
\caption{{\it Left panel:} The optical photon counts of each burst is plotted here against the X-ray photon counts. The straight line represents the best fit through the origin. {\it Right panel:} The optical conversion factor of each burst is plotted here against the X-ray photon counts.}
\end{figure}

\begin{figure}
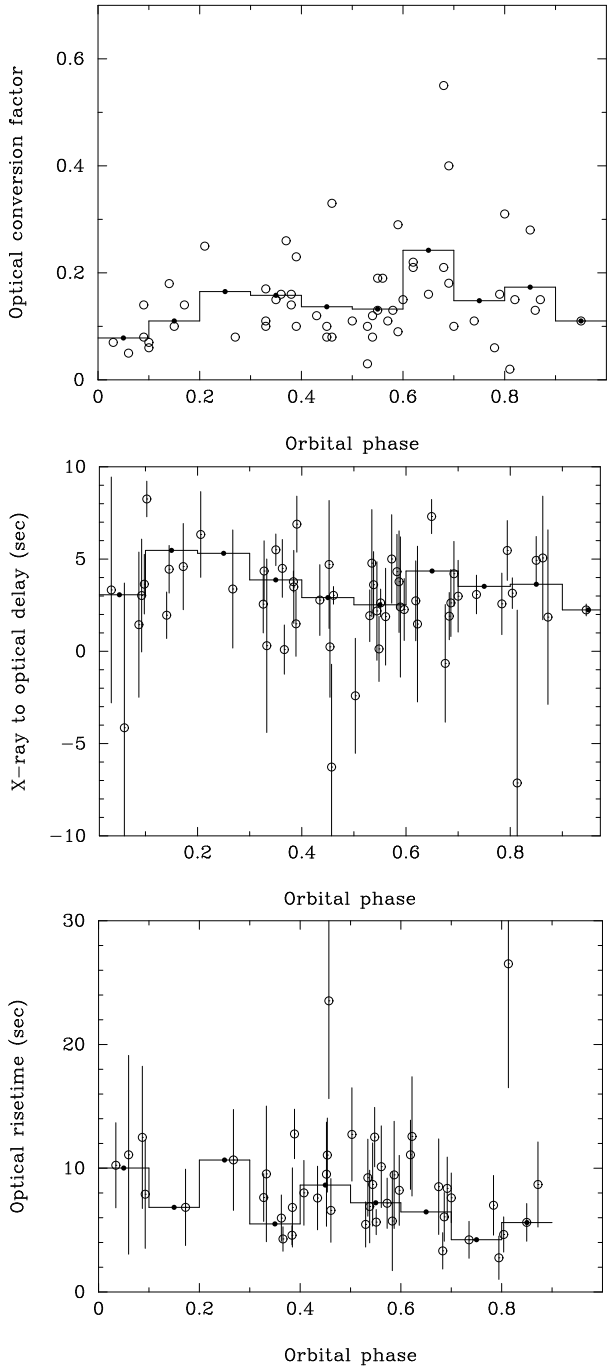

\includegraphics[width=60mm,angle=-90]{Fig5a.ps}\\
\includegraphics[width=60mm,angle=-90]{Fig5b.ps}\\
\includegraphics[width=60mm,angle=-90]{Fig5c.ps}
\caption{{\it Top panel:} The Optical conversion factor of each burst is plotted against the orbital phase.
{\it Middle panel:} The delay between the X-ray and optical start time of each burst is plotted here against the orbital phase. {\it Bottom panel:} The rise time of the optical burst is plotted here against the orbital phase.
In each panel, the lines plotted through the filled cirles show the values averaged over every orbital phase bin of 0.1.}
\end{figure}
\end{document}